\documentclass[a4paper,11pt]{article}
\usepackage{pos}
\usepackage[export]{adjustbox}

\def\powhegbox{\textsc{Powheg-Box}}
\def\pythia{\textsc{Pythia8}}
\def\mg5{\textsc{MG5\_aMC@NLO}}

\def\sherpa{\textsc{Sherpa}}

\title{Parton shower effects in $t\bar{t}W^\pm$ production at NLO QCD}

\author*[a]{Manfred Kraus}

\affiliation[a]{Physics Department, Florida State University, Tallahassee, FL 32306-4350, U.S.A}

\emailAdd{mkraus@hep.fsu.edu}

\abstract{
We present a selection of results from our recent study of $pp\to t\bar{t}W^\pm$ 
production matched to partons showers at NLO QCD at the LHC. Theoretical 
predictions are obtained at perturbative orders $\mathcal{O}(\alpha_s^3\alpha)$ 
and $\mathcal{O}(\alpha_s\alpha^3)$, where the different contributions are 
studied first separately at the inclusive level before being combined within a
realistic two same-sign lepton signature. We investigate in detail uncertainties 
originating from missing higher-order corrections and from the parton-shower 
matching scheme employed.
}

\FullConference{%
  *** The European Physical Society Conference on High Energy Physics (EPS-HEP2021), ***\\
  *** 26-30 July 2021 ***\\
  *** Online conference, jointly organized by Universität Hamburg and the research center DESY ***
}

\begin{document}
\maketitle

\section{Introduction}
Top-quark pair production in association with a $W$ gauge boson is one of the 
rarest scattering processes in the Standard Model (SM). Nonetheless, it has 
received much attention recently as it constitutes a large background to many SM
measurements and searches for physics beyond the Standard Model (BSM).
Most notably, the $pp\to t\bar{t}W^\pm$ process is the dominant background for
SM measurements of the $t\bar{t}H$ and of the four top-quark production process 
in the multi-lepton signatures. However, in both cases, recent measurements of 
the $t\bar{t}W$ background contributions reveal tensions with the SM 
predictions~\cite{ATLAS:2019nvo}.

In recent years a lot of progress has been made to improve the theoretical
description of the $pp\to t\bar{t}W^\pm$ process. For instance, NLO QCD and 
electroweak (EW) corrections for predictions based on stable top-quarks are 
already known for some time~\cite{Maltoni:2015ena,Frixione:2015zaa,
Frederix:2017wme,Frederix:2018nkq}. Furthermore, the impact of threshold 
resummation on total cross sections and differential distributions has been 
studied in Refs.~\cite{Li:2014ula,Broggio:2016zgg,Kulesza:2018tqz,
Broggio:2019ewu,Kulesza:2020nfh}. In addition, also the NLO QCD corrections to 
the decay in the NWA has been studied~\cite{Campbell:2012dh} as well as the 
inclusion of off-shell effects and non-resonant contributions has been addressed 
for the first time in Refs.~\cite{Bevilacqua:2020srb,Bevilacqua:2020pzy,
Denner:2020hgg} for the dominant QCD production mode as well as for the full 
one-loop SM corrections in Ref.~\cite{Denner:2021hqi}. Furthermore, the process 
has been matched to parton showers~\cite{Garzelli:2012bn,Maltoni:2014zpa,
Cordero:2021iau} and effects from multi-jet merging have been 
studied~\cite{Frederix:2020jzp,vonBuddenbrock:2020ter,Frederix:2021agh} as well. 
Also the approximate inclusion of full off-shell effects in parton-shower matched
calculations of on-shell $t\bar{t}W$ has been investigated in 
Ref.~\cite{Bevilacqua:2021tzp}.

In the following, we present results for theoretical predictions for the 
$pp\to t\bar{t}W^\pm$ process at $\mathcal{O}(\alpha_s^3\alpha)$ and
$\mathcal{O}(\alpha_s\alpha^3)$ matched to parton showers via the \powhegbox{}
framework.

\section{Outline of the calculation}
%
\begin{figure}[ht!]
 \centering
 \includegraphics[width=\textwidth]{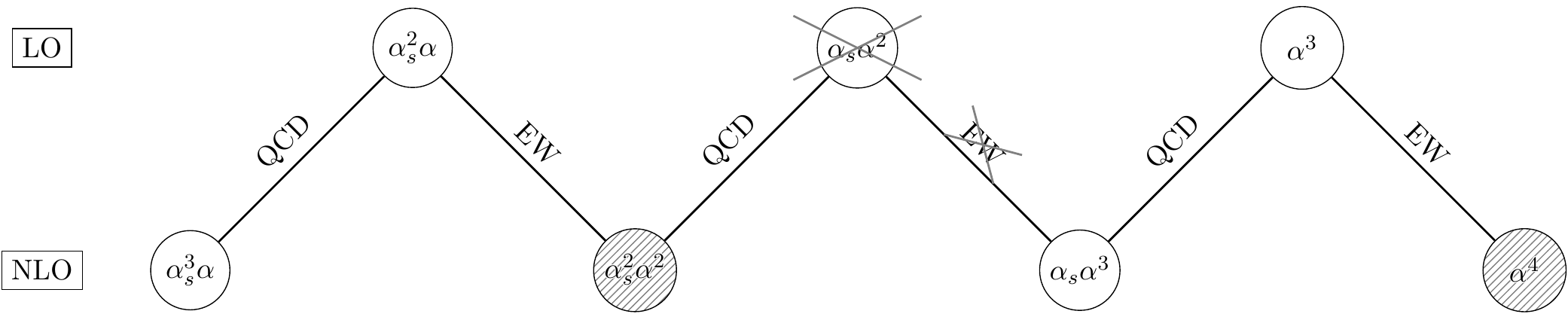}
 \caption{The structure of higher-order corrections for the production of the 
 $pp\to t\bar{t}W^\pm$ final state.}
 \label{fig:coup}
\end{figure}
In Fig.~\ref{fig:coup} the different perturbative orders that contribute to the
$pp\to t\bar{t}W^\pm$ process up to the one-loop level are depicted. At the 
leading order only the order $\mathcal{O}(\alpha_s^2\alpha)$ and 
$\mathcal{O}(\alpha^3)$ contribute, which correspond to the squared QCD and EW
born matrix elements. The interference at $\mathcal{O}(\alpha_s\alpha^2)$ vanishes
due to color conservation, which allows to identify at the next-to-leading order 
the terms at $\mathcal{O}(\alpha_s^3\alpha)$ and $\mathcal{O}\alpha_s\alpha^3)$ 
as pure QCD corrections, while the contributions at 
$\mathcal{O}(\alpha_s^2\alpha^2)$ are mixed QCD-EW and $\mathcal{O}(\alpha^4)$ 
are pure EW corrections.

Our calculation takes into account the dominant higher-order corrections at the
perturbative orders $\mathcal{O}(\alpha_s^3\alpha)$ and
$\mathcal{O}(\alpha_s\alpha^3)$. The implementation of the dominant QCD 
corrections in the \powhegbox{} is based on virtual amplitudes provided via 
\textsc{NLOX}~\cite{Honeywell:2018fcl,Figueroa:2021txg}. We compare results 
obtained with the \powhegbox{} to those obtained with \mg5{}~\cite{Alwall:2014hca}
as well as the \sherpa{} framework~\cite{Gleisberg:2008ta,Sherpa:2019gpd}. 
Spin-correlated top-quark decays are, depending on the framework employed, taken 
into account in the study of the two same-sign lepton signature according to 
Refs.~\cite{Frixione:2007zp,Artoisenet:2012st,Richardson:2001df}. Theoretical 
predictions based on the \powhegbox{} and \mg5{} are interfaced with the 
\pythia{} parton shower, while for \sherpa{} its own Catani-Seymour shower is 
used. Thus, our comparison allows to compare different parton-shower matching 
schemes as well as different parton showers. Further details on the specific 
setup of the comparison can be found in Ref.~\cite{Cordero:2021iau}.

\section{Inclusive $t\bar{t}W$ production}
First we compare the different Monte Carlo generators at the fully inclusive
level with stable top quarks and $W$ gauge bosons. Jets are defined via the
anti-$k_T$ jet algorithm with a separation parameter of $R=0.4$ and we require
jets to fulfill the following constraints
\begin{equation}
 p_T(j) > 25~\text{GeV}\;, \qquad |y(j)| < 2.5\;, \qquad N_\textrm{jets} \geq 0\;.
\end{equation}

\begin{figure}[ht!]
 \centering
 \includegraphics[width=0.49\textwidth]{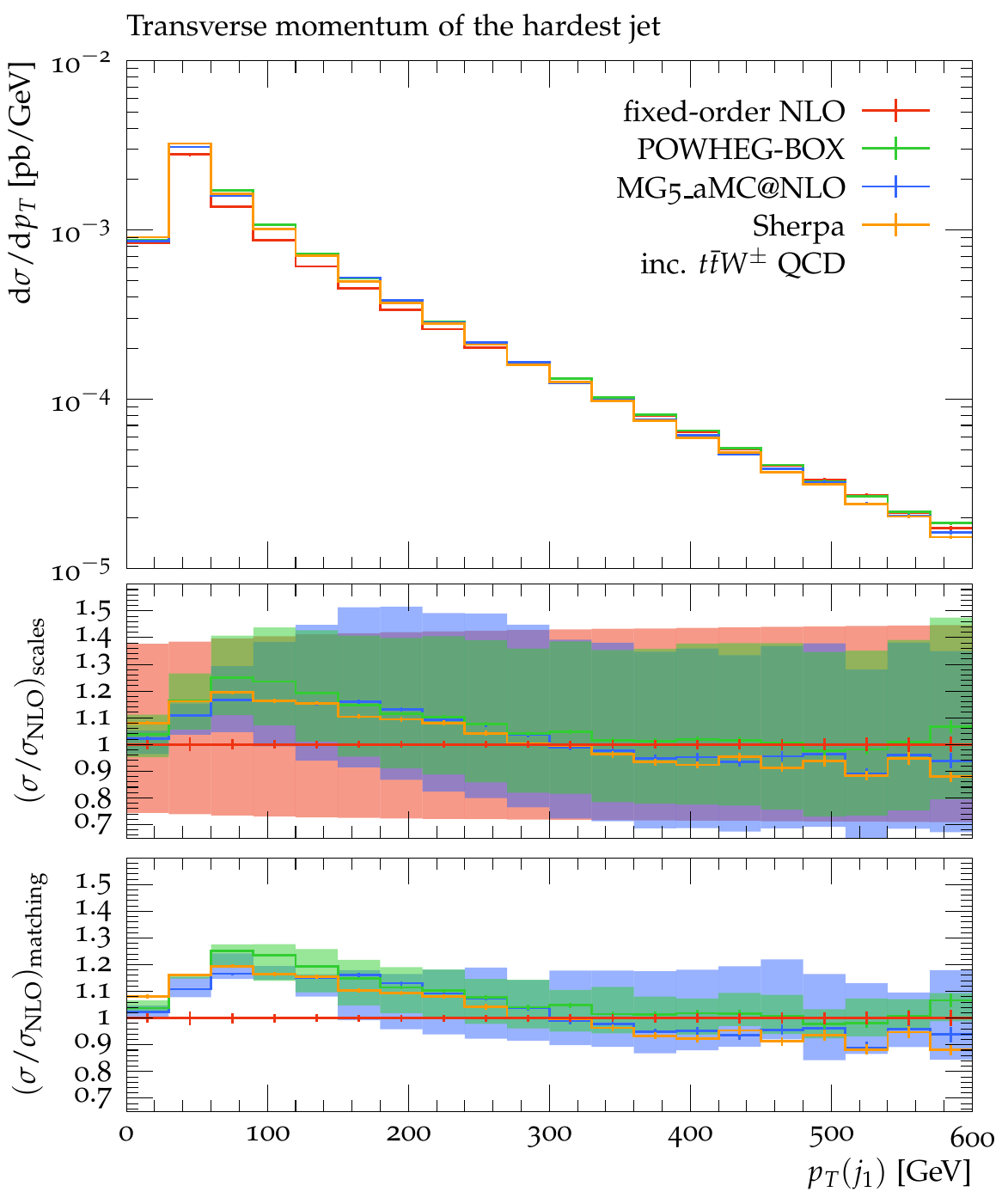}
 \includegraphics[width=0.49\textwidth]{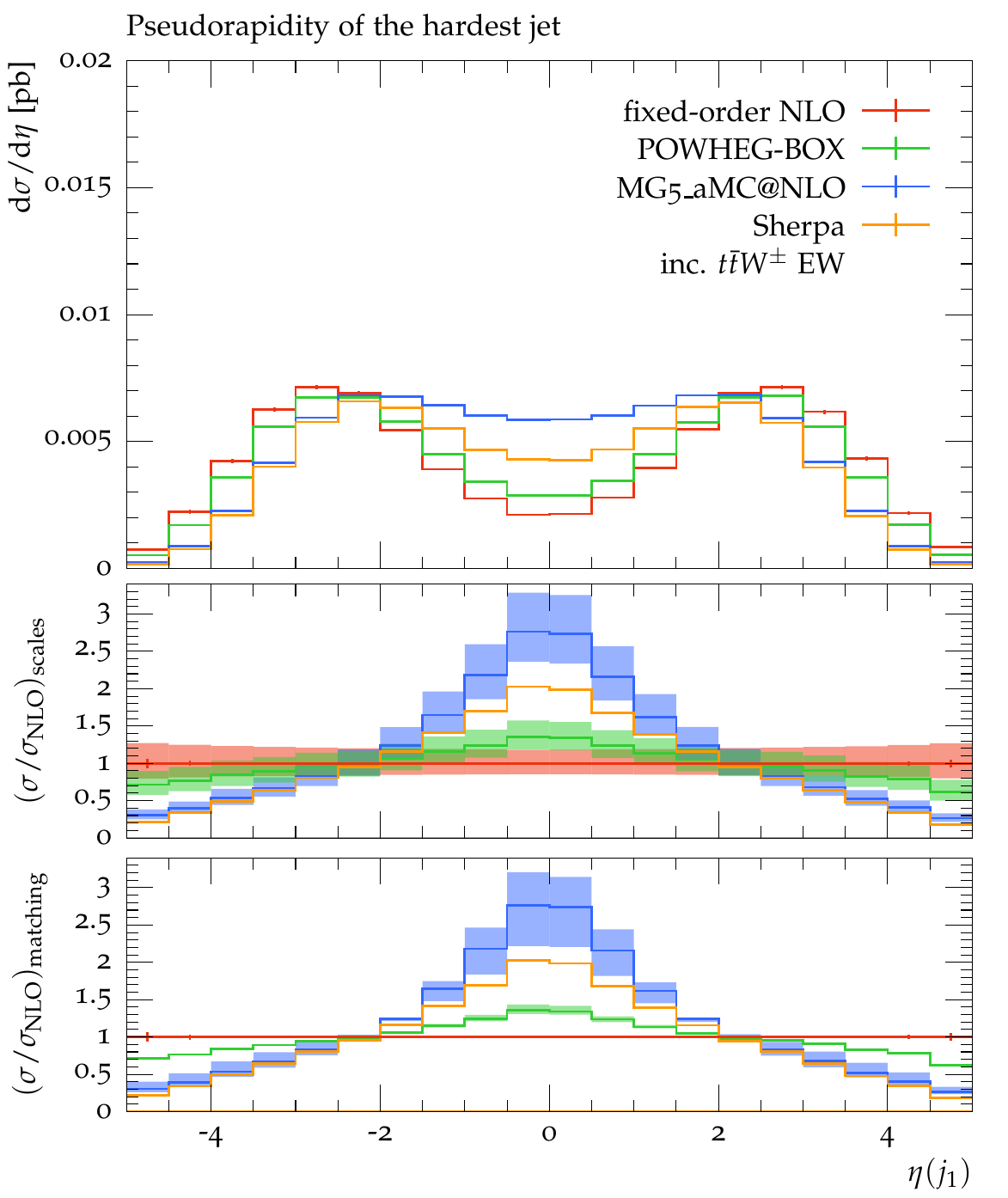}
 \caption{Differential cross section distribution as a function of the transverse 
momentum of the hardest jet at $\mathcal{O}(\alpha_s^3\alpha)$ (l.h.s) and of
the rapidity of the hardest jet at $\mathcal{O}(\alpha_s\alpha^3)$ (r.h.s).}
 \label{fig:inc}
\end{figure}
In Fig.~\ref{fig:inc} we highlight the transverse momentum of the leading jet for
the QCD production mode of $pp\to t\bar{t}W^\pm$ at 
$\mathcal{O}(\alpha_s^3\alpha)$ and the rapidity of the leading jet for the EW 
production channel at $\mathcal{O}(\alpha_s\alpha^3)$. Even though the presented
observables have only leading-order accuracy they serve as a good representation 
of our findings also for inclusive NLO accurate observables. We refer to 
Ref.~\cite{Cordero:2021iau}, where we have studied more distributions for both 
production channels.
For the transverse momentum distribution we find very good agreement between all
employed generators. All predictions including parton-shower effects align very 
well with the fixed-order NLO QCD result in the tail of the distribution. This is
expected, since the hard matrix elements are reliable for hard emissions and 
shower corrections are small in this region. On the other hand, we find 
differences between shower based and fixed-order predictions at the level of 
$20\%$ at the beginning of the spectrum, which is dominated by soft and collinear
emissions. These differences can be attributed to the leading logarithmic 
resummation performed by the parton showers. Missing higher-order corrections 
dominate the uncertainties for most of the plotted range. Only at the beginning 
of the distribution matching uncertainties become comparable in size. 
On the contrary, we observe large differences in the EW predictions at 
$\mathcal{O}(\alpha_s\alpha^3)$ as can be seen in the rapidity distribution on 
the right of Fig.~\ref{fig:inc}. None of the parton-shower based predictions is 
in agreement with fixed-order calculation. To be specific, we find deviations of 
the order of $50\%$ for the \powhegbox{}, $100\%$ for \sherpa{} and more than 
$200\%$ for \mg5{}. As indicated by the large matching uncertainties of the 
\mg5{} prediction the resulting curve depends crucially on the choice of the 
initial shower scale.
 
\section{Two same-sign lepton signature}
Turning now to the comparison for a realistic two same-sign lepton signature at 
the fiducial level. Final-state particles are subject to the following phase 
space cuts
\begin{equation}
\begin{split}
 &p_T(\ell) > 15~\text{GeV}\;, \qquad |\eta(\ell)| < 2.5\;, \qquad 
 p_T(j) > 25~\text{GeV}\;, \qquad |\eta(j)| < 2.5\;, 
\end{split}
\end{equation}
where jets are formed using the anti-$k_T$ jet algorithm with $R=0.4$. 
Additionally, we require exactly 2 same-sign leptons, at least $2b$ jets and 
at least $2$ light jets. Predictions shown in the following correspond to the
sum of $t\bar{t}W^+$ and $t\bar{t}W^-$ processes as well as the combination of
$\mathcal{O}(\alpha_s^3\alpha)$ and $\mathcal{O}(\alpha_s\alpha^3)$ contributions.

\begin{figure}[ht!]
 \centering
 \includegraphics[valign=t,width=0.49\textwidth]{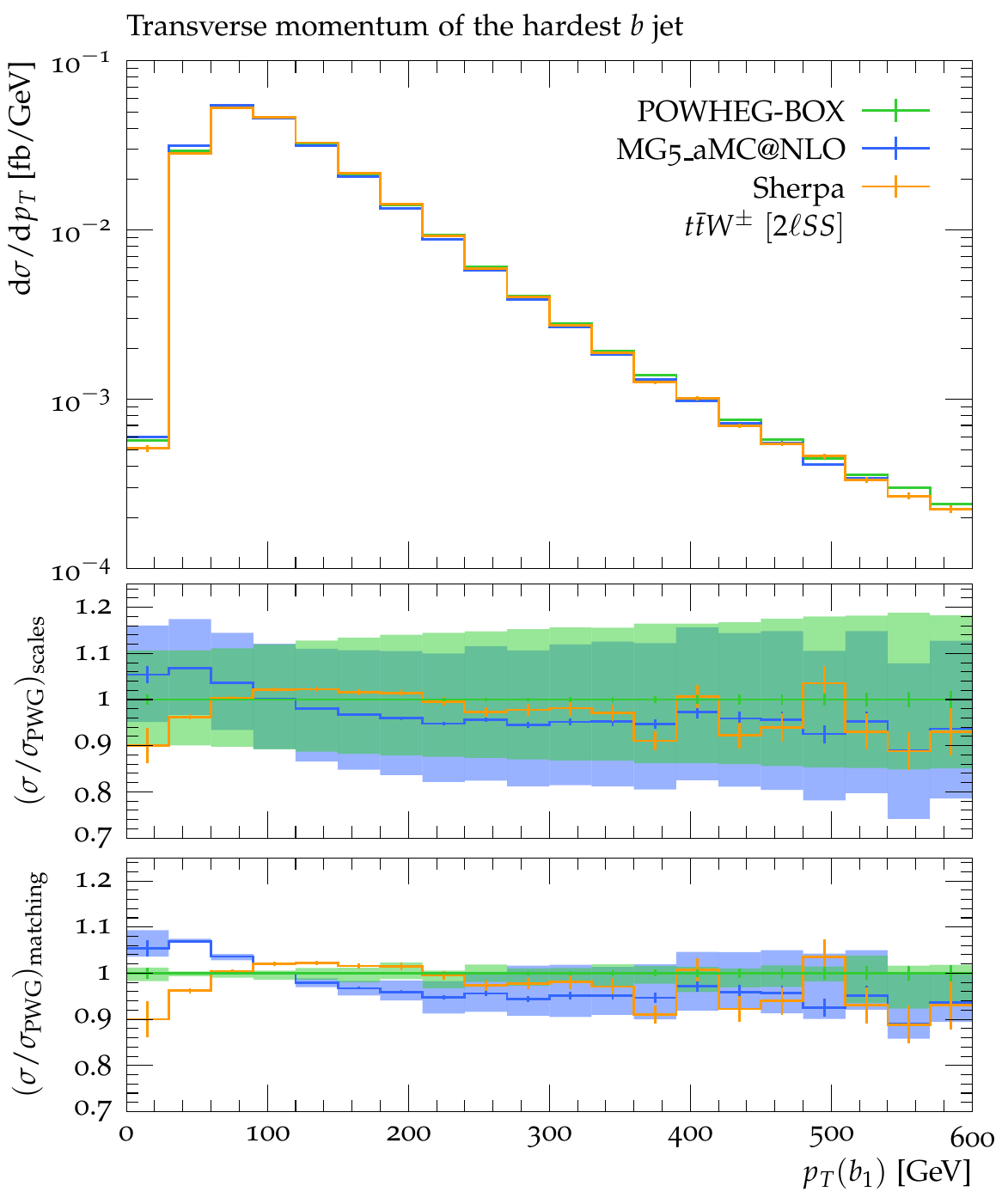}
 \includegraphics[valign=t,width=0.49\textwidth]{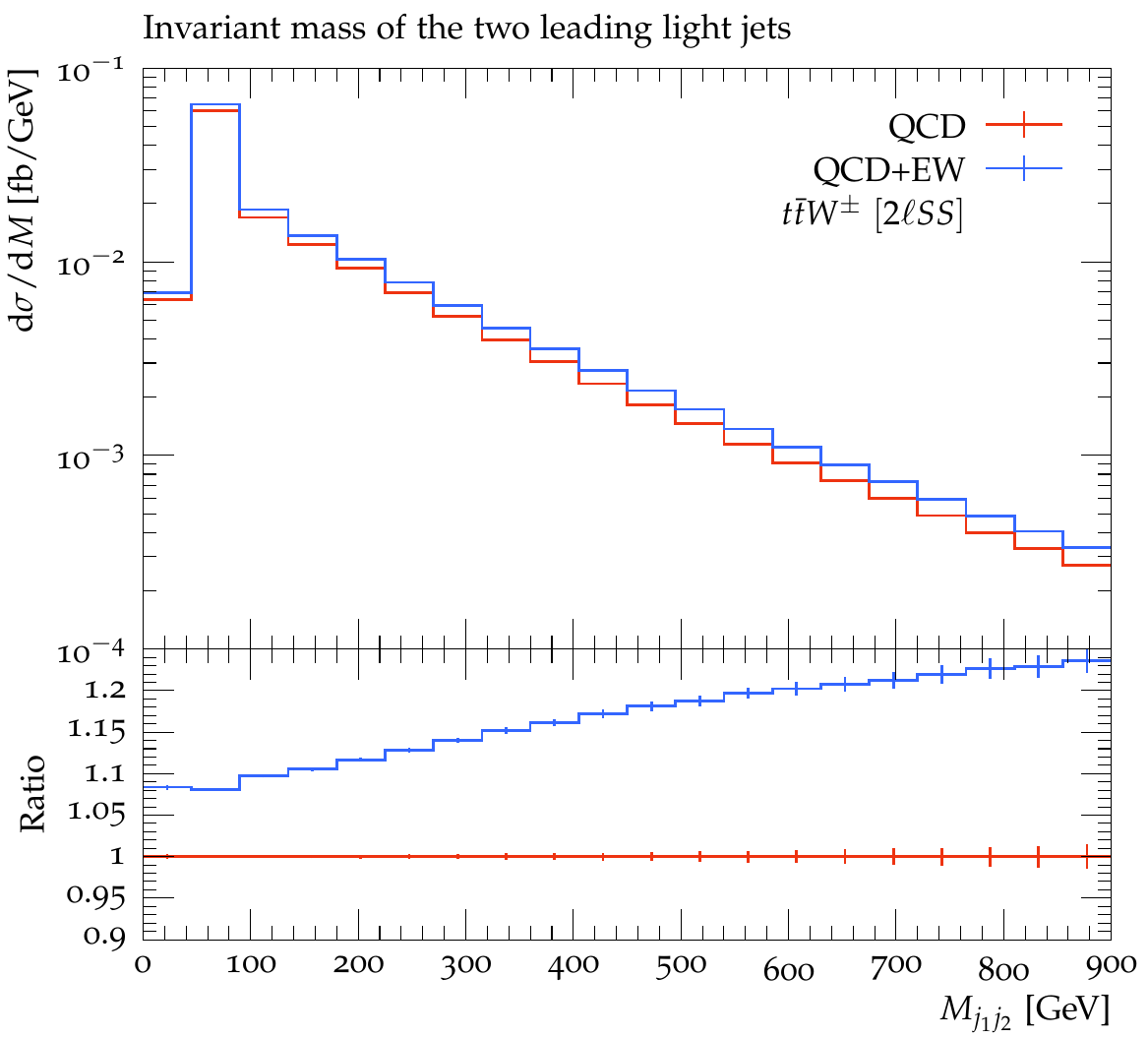}
 \caption{Differential cross section distribution as a function of the transverse 
momentum of the hardest $b$ jet (l.h.s) and of the invariant mass of the two 
hardest light jets (r.h.s).}
\label{fig:fid}
\end{figure}
Representative of our full findings in Ref.~\cite{Cordero:2021iau} we show the
transverse momentum distribution of the leading $b$ jet on the left and the 
invariant mass distribution of the two hardest light jets on the right of 
Fig.~\ref{fig:fid}. The hardest $b$ jet predominantly originates from the 
top-quark decay and therefore allows for a comparison of the different approaches
to model spin-correlated decays in the event generators. We find very good 
agreement between the various predictions with differences less than $5\%$ for 
most of the plotted range. In the beginning of the distribution differences are 
slightly larger at the level of $10\%$. This, however, can be attributed to the 
different treatment of radiation from heavy quarks. Over the whole plotted range 
we also find that scale uncertainties dominate the theoretical uncertainties and 
are of the order of $10\%-20\%$.
From the invariant mass distribution of the two hardest light jets, as depicted 
on the right of Fig.~\ref{fig:fid}, we can draw multiple conclusions. First of 
all, we notice the that the peak of the distribution is located around the $W$ 
boson resonance, i.e. the leading jets originate from the hadronic decaying $W$ 
boson that is described at leading-order accuracy for all generators. 
Furthermore, we see that the EW production mode starts as a $+10\%$ correction 
that increases towards the end of the spectrum up to $+25\%$. Generally, the EW
contribution becomes sizable if the considered observable is sensitive to forward
jets. However, for most observables we studied the inclusion of the 
$\mathcal{O}(\alpha_s\alpha^3)$ contribution constitutes a constant $+10\%$ 
correction at the differential level.

\section{Summary}
We presented results of our recent comparison~\cite{Cordero:2021iau} of the 
$pp\to t\bar{t}W^\pm$ process for a two same-sign lepton signature. We find 
overall very good agreement between the various generators employed in the case 
of the QCD production mode at $\mathcal{O}(\alpha_s^3\alpha)$. On the contrary, 
for the EW production of the $pp\to t\bar{t}W^\pm$ final state we observe sizable
differences between the generators. Nonetheless, the less accurate modeling of 
the EW contribution has only a small impact once QCD and EW contributions are
combined because the latter are generally a $10\%$ effect on top of the dominant
QCD contribution. Only in phase space regions dominated by forward jets the 
EW contributions becomes sizable. Furthermore, we investigated spin-correlation 
effects in the top-quark decay modeling, which can also modify the shape of  
leptonic observables at the $10\%$ level. 

Further improvements in the realistic description is highly signature dependent. 
For instance, in multi-lepton signatures the NNLO QCD corrections to the 
$pp\to t\bar{t}W^\pm$ production process are of utmost importance. For 
multi-lepton signatures also the matching of the full off-shell computation to 
parton showers will further improve the description of fiducial phase space 
volumes. However, for signatures involving the hadronic $W$ boson decays the 
inclusion of NLO QCD corrections in the decay is inevitable.

\subsection*{Acknowledgements}

The author acknowledges support by the U.S. Department of Energy under the grant
DE-SC0010102.

\end{document}